\documentclass[aps,pra,superscriptaddress,preprint,showpacs,amsmath,amsfonts,amssymb]{revtex4-1}
\usepackage{amsmath,amsfonts,amssymb,color}
\usepackage{amsthm}
\usepackage{leftidx}
\usepackage{graphicx}
\usepackage{xcolor}
\usepackage{dcolumn}
\usepackage{bm}
\usepackage{epstopdf}
\usepackage{epsfig}

\newcommand{\n}{\noindent}

\begin{document}
	\title{Nonlinear Dirac cones}
	\author{Raditya Weda Bomantara}
	\affiliation{%
		Department of Physics, National University of Singapore, Singapore 117543
	}
	
	\author{Wenlei Zhao}
	\affiliation{%
		 School of Science, Jiangxi University of Science and Technology, Ganzhou 341000, China
	}
	\affiliation{%
		Department of Physics, National University of Singapore, Singapore 117543
	}

	\author{Longwen Zhou}
	\email{zhoulw13@u.nus.edu}
	\affiliation{%
		Department of Physics, National University of Singapore, Singapore 117543
	}
	\author{Jiangbin Gong}%
	\email{phygj@nus.edu.sg}
	\affiliation{%
		Department of Physics, National University of Singapore, Singapore 117543
	}
	\date{\today}
	

   	\vspace{2cm}

	\begin{abstract}
	Physics arising from two-dimensional~($2$D) Dirac cones has been a topic of great theoretical and experimental interest to studies of gapless topological phases and to simulations of relativistic systems. Such $2$D Dirac cones are often characterized by a $\pi$ Berry phase and are destroyed by a perturbative mass term. By considering mean-field nonlinearity in a minimal two-band 
	Chern insulator model, we obtain a novel type of Dirac cones that are robust to local perturbations without symmetry restrictions. Due to a different pseudo-spin texture,
		the Berry phase of the Dirac cone  is no longer quantized in $\pi$, and can be continuously tuned as an order parameter. Furthermore, in an Aharonov-Bohm~(AB) interference setup to detect such Dirac cones, the adiabatic AB phase is found to be $\pi$ both theoretically and computationally, offering an observable topological invariant and a fascinating example where the Berry phase and AB phase are fundamentally different.
	We hence discover a nonlinearity-induced quantum phase transition from a known topological insulating phase to an unusual gapless topological phase.

	\end{abstract}


	\maketitle
	
	\textit{Introduction.--} Starting from the seminal papers by Thouless et al.~\cite{Thou1,Thou2}, the role of topology in band theory of solids has attracted tremendous interest. In addition to topological insulators~\cite{TI1,TI2,TI3}, the search for novel topological materials has led to discoveries of Dirac~\cite{DSM01,DSM02,DSM03,DSM1,DSM2,DSM3}, Weyl~\cite{WSM1,Burkov,WSM2,WSM3}, and nodal line semimetals~\cite{Burkov,Heikila,linesm1,linesm2,linesm3,linesm4}. 
	Recently, topological phases in interacting systems have stimulated much attention~\cite{inttp1,inttp2,inttp3,inttp4,inttp5,inttp6,inttp7}.
In particular, after the topological classification of noninteracting topological insulators~\cite{class1,class2}, a general topological classification of interacting systems constitutes an important topic~\cite{inttp1,inttp4,inttp7}.

 	Developing physical insights into the interplay of topology and interaction, which typically requires the use of advanced
 		many-body techniques~\cite{TopoIntBook}, is often a challenge.  On the other hand,  simple mean-field approaches may be still fruitful \cite{mean1,mean2,mean3,mean4,mean5,mean6}.  Here we take a modest mean-field approach to a minimal two-band topological insulator model with on-site bosonic interactions. This leads to a nonlinear problem, insofar as the Bloch states are now eigenstates of the Gross-Pitaevskii~(GP) equation~\cite{Gross,Pitaevskii} depicting a two-dimensional~($2$D) tight-binding lattice with on-site mean-field potential. GP equations are a well-known tool to study Bose-Einstein condensate~(BEC) of cold atomic gases, such as matter-wave solitons~\cite{sol1,sol2,sol3}. Moreover, GP equations are also useful in the study of photonic systems,
 	where Kerr nonlinearity becomes important with the increase of light intensity~\cite{kerr1,mean3,kerr3}.
    We expect our theoretical predictions below to be relevant to simulations of topological quantum matter in cold atom and photonic systems~\cite{Jotzu2014,Goldman2016,Lu2014}.
 	
   	As already learned from zero- or one-dimensional systems, the band structure arising from solving the stationary GP equation  may accommodate self-crossing loop~(swallowtail) formations~\cite{loop1}. This feature in $2$D situations suggests the loss of a well-defined band Chern number as a topological invariant. In the vicinity of the self-crossing point of a $2$D looped band, we discover the formation of a novel type of $2$D Dirac cones. Such nonlinear Dirac cones~(NDCs) share analogous robustness with Weyl points in Weyl semimetals \cite{WSM1,Burkov,WSM2,WSM3}.
   Equally interesting, due to a peculiar pseudo-spin texture different from what is found in those familiar Dirac cones in $2$D Dirac semimetals~\cite{DSM01,DSM02,DSM03,DSM1,DSM2,DSM3}, NDCs yield Berry phases no longer quantized in $\pi$. Further,
    their band structure may be potentially useful for quantum simulation of some exotic physics.

 	Following the stimulating experiment reported in Ref.~\cite{ABexp}, we propose to detect the formation of an NDC by use of an interference setup in the spirit of the Aharonov-Bohm~(AB) effect \cite{ABphase}.
 As a remarkable finding detailed below, the AB phase associated with two adiabatic paths in the reciprocal space enclosing an NDC is still quantized in $\pi$. This provides a stimulating example where the Berry phase and AB phase are different. We shall also use this result to identify an experimentally accessible topological invariant for NDCs.

 	\textit{Two-band Model.--} Consider a nonlinear version of the spinless (bosonic) two-band Dirac-Chern insulator model~\cite{QWZ}, with two sublattices serving as pseudo-spin. With the lattice Hamiltonian detailed in Supplementary Material \cite{supp}, the stationary GP equation in the momentum space assumes the following form,
 	\begin{equation}
 	\mathcal{H}[k_x,k_y,\psi(k_x,k_y)]|\psi(k_x,k_y)\rangle=E(k_x, k_y)|\psi(k_x,k_y)\rangle,
 	\label{GPeq1}
 	\end{equation}
 	with
	\begin{equation}
	\mathcal{H}[k_x,k_y,\psi(k_x,k_y)]= J_x\sin(k_x)\sigma_x+J_y\sin(k_y)\sigma_y+\mathcal{B}(k_x,k_y)\sigma_z +g\left[\begin{array}{cc} |\psi_1(k_x,k_y)|^2 & 0	\\ 0 & |\psi_2(k_x,k_y)|^2 \end{array}\right] \;,
	\label{nlqwz2}
	\end{equation}
	\n where $\sigma$'s are Pauli matrices in the usual representation, $k_x$~($k_y$) is the quasimomentum along $x$~($y$) direction, $|\psi(k_x,k_y)\rangle\equiv [\psi_1(k_x,k_y), \psi_2(k_x,k_y)]^{T}$ denotes a Bloch band state with two pseudo-spinor components, $g$ is the nonlinear strength, $\mathcal{B}(k_x,k_y)=B[M+\cos(k_x)+\cos(k_y)]$, $B$ is a parameter that may be interpreted as the hopping strength, $M$ is related to the difference in potential strength between the two lattice species, $J_{x}$ and $J_{y}$ mimic the effect of Rashba-like spin-orbit couplings. Throughout this paper, all physical variables are assumed to be scaled and hence in dimensionless units. The linear version of the model here shares the same topological property with the Haldane model already realized in cold-atom systems~\cite{Jotzu2014}, with its spinful counterpart~\cite{BernevigScience} describing a quantum spin Hall insulator also realized experimentally~\cite{BHZScience}. Therefore, $\mathcal{H}[k_x,k_y,\psi(k_x,k_y)]$ is a representative and natural choice to accommodate bosonic mean-field interactions and to motivate experimental investigations.

	\begin{figure}
		\begin{center}
			\includegraphics[scale=0.52]{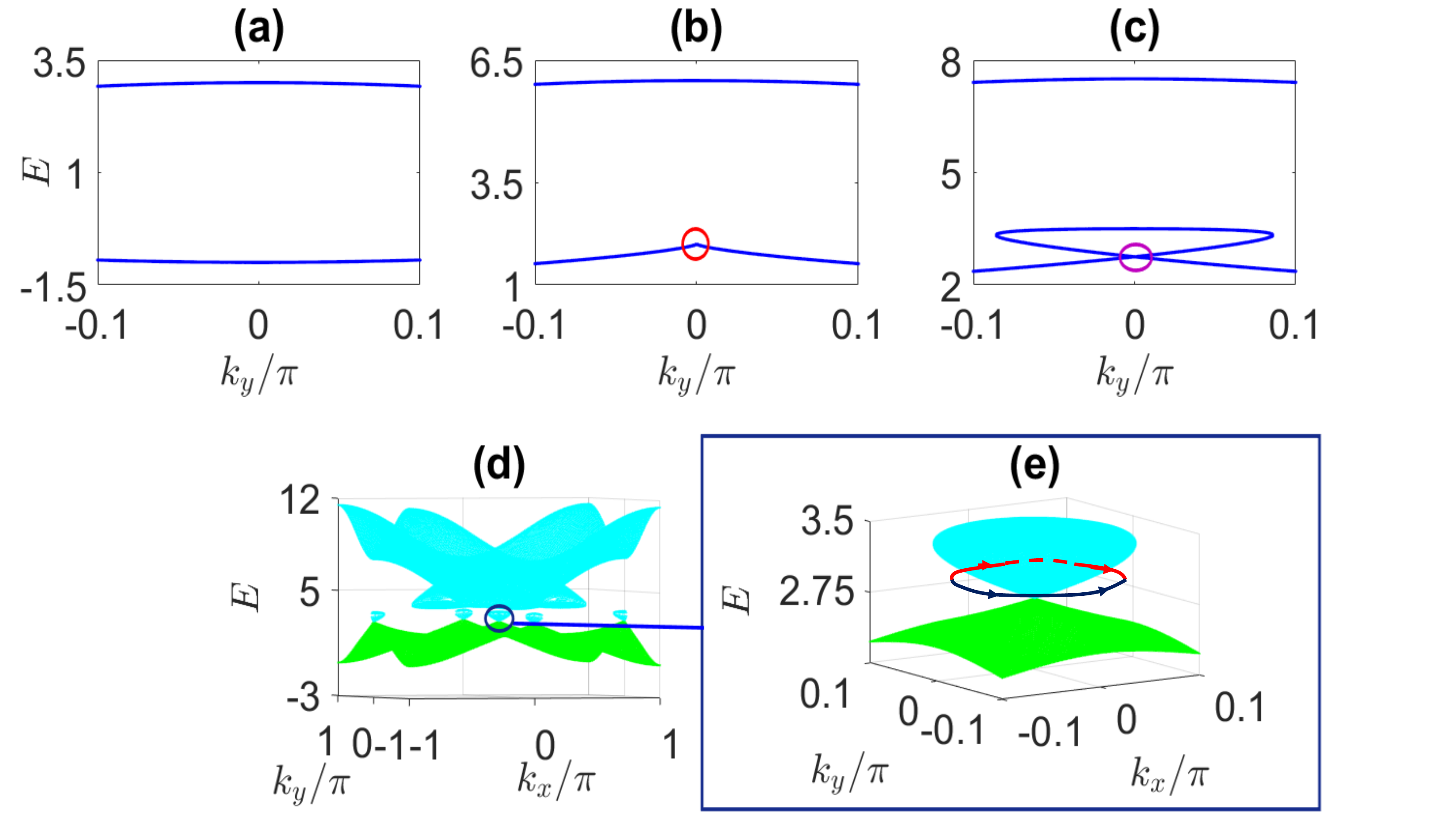}
		\end{center}
		\caption{(color online). The development of loop structure in the nonlinear Dirac-Chern insulator model. The parameters chosen are $M=-1$, $B=2$, $J_x=J_y=1$, $k_x=0$, (a) $g=1$, (b) $g=4$, and (c) $g=5.5$. The full spectrum over the 2D Brillouin zone is plotted in (d). Panel (e) shows an enlarged version of (d) near a looped band structure. The red and blue curves in (e) illustrate two different interfering paths in a proposed AB-effect experiment.}
		\label{fnlQWZ}
	\end{figure}

 	Figure~\ref{fnlQWZ}(a)-(c) show the snapshots of the band structure near $k_y=0$ at a fixed $k_x=0$. As expected from previous theoretical and experimental studies of nonlinear Bloch bands in zero or one dimensional systems~\cite{loop1,loops,loop2,loop3,loopliu2,loop4,loop5,loops2,loopgong}, a self-crossing loop structure emerges as $g$ increases beyond a critical value $g_c$. Specifically, for $g=g_c$, the bottom band starts to develop a cusp~[see the red circle in Fig.~\ref{fnlQWZ}(b)]. For $g>g_c$,  this cusp transforms into a self-crossing loop, with the self-crossing point marked by the magenta circle in Fig.~\ref{fnlQWZ}(c). Next, we examine in Fig.~\ref{fnlQWZ}(d) the complete band structure for the whole $2$D Brillouin zone (BZ), with special attention paid to the bottom band.  There, the self-crossing loop structure is found to form along both $k_x$ and $k_y$ dimensions. In particular, three looped subbands have grown from the bottom mother band (two of them at the edge of the shown BZ). Within the regime of each individual looped structure, Eq.~(\ref{nlqwz2}) yields four Bloch states. Precisely at a self-crossing point of a looped subband, two of the four Bloch states are degenerate.
 	The Chern number of the lowest band, which distinguish between topologically trivial and nontrivial phases in noninteracting Dirac-Chern insulators, stays at the same value as its linear counterpart~($g=0$) until $g=g_c$. For $g>g_c$, though the bottom mother band is still well separated from the upper band,  its Chern number becomes ill defined due to the emergence of the self-crossing points. As such, it is necessary to study the bottom mother band individually from the perspective of a gapless topological phase.

	\textit{Nonlinear Dirac cones.--} For the rest of our analysis, we set $J_x=J_y=1$ and let $\mathcal{B}\equiv \mathcal{B}(0,0)=B(M+2)$. A simple calculation indicates that $g_c=2\mathcal{B}$, beyond which Eq.~(\ref{nlqwz2}) may host four stationary solutions for one given set of $k_x$ and $k_y$.
	By considering Eq.~(\ref{GPeq1}) near one of the three self-crossing points seen in Fig.~\ref{fnlQWZ}(d), e.g., at $k_x=k_y=0$, we obtain the following energy solutions~\cite{supp},
	\begin{equation}
	E_{\pm}(k_x,k_y) = \frac{g}{2}\pm \frac{1}{\sqrt{1-\frac{4\mathcal{B}^2}{g^2}}}\sqrt{k_x^2+k_y^2}\;,
	\label{loop}
	\end{equation}
	where $+$~($-$) stands for the upper~(lower) energy branch around a self-crossing point. These energy solutions are isotropic in the $k_x$-$k_y$ plane, linear with respect to the magnitude of the overall wavevector $k\equiv \sqrt{k_x^2+k_y^2}$. Analogous results with linear energy solutions in both $k_x$ and $k_y$ are found near other self-crossing points. We thus witness here the emergence of  $2$D self-crossing perfect Dirac cones for $g>g_c$, which we will refer to as nonlinear Dirac cones (NDCs).

	To develop further insights, we rewrite Eq.~(\ref{nlqwz2}) by making use of the above-obtained energy solutions and its associated stationary states \cite{supp}, arriving at two effective Hamiltonians for the positive and negative branches of the Dirac cone centered at $k_x=k_y=0$,
	\begin{equation}
	h_{\rm eff,\pm}= k_x\sigma_x +k_y \sigma_y\mp \frac{2\mathcal{B}}{\sqrt{g^2-4\mathcal{B}^2}}\sqrt{k_x^2+k_y^2} \sigma_z \;.
	\label{nleffh}
	\end{equation}
 	\n That the two branches of the Dirac cone are described by different effective Hamiltonians is simply because $\mathcal{H}$ specified in Eq.~(\ref{nlqwz2}) depends on the Bloch band state. As a consequence, the two Bloch band states associated with the positive and negative branches of the same Dirac cone are not orthogonal in general.
 	Of particular importance and interest, $h_{\rm eff,\pm}$ differ from the familiar effective Hamiltonian of a conventional $2$D Dirac cone~\cite{DSM1,DSM2}, in that only $h_{\rm eff,\pm}$ found here has a $\sigma_z$ (mass) term.
 To our knowledge, the peculiar form of
 	$h_{\rm eff,\pm}$ depicted in Eq.~(\ref{nleffh}) is not obtained previously in any condensed matter system. Three implications are detailed below.

 	Firstly, NDCs are robust against generic local perturbations, similar to Weyl points in Weyl semimetals~\cite{WSM1,Burkov,WSM2,WSM3}. To understand this, note that a generic perturbation in two-level systems can be expressed in terms of the Pauli matrices $\sigma_x$, $\sigma_y$, and $\sigma_z$. As seen from $h_{\rm eff,\pm}$, perturbations proportional to $\sigma_x$ or $\sigma_y$ will only shift the location of the self-crossing point of an NDC in the $k_x$-$k_y$ plane, {\it e.g.} perturbation of the form $h_x\sigma_x$ shifts the location of the self-crossing point from $(k_x,k_y)=(0,0)$ to $(k_x,k_y)=(-h_x,0)$. For a perturbation proportional to $\sigma_z$, {\it i.e.}, a perturbative mass term (which opens a gap in conventional 2D Dirac cones), it can only renormalize the value of $\mathcal{B}$ \cite{footnote}.
 	This preserves NDCs again so long as $g$ is not too close to $g_c$.  These understandings are computationally confirmed in Supplementary Material ~\cite{supp}.
 	 While a conventional $2$D Dirac cone needs to be protected by certain symmetries, NDCs here are protected by interaction.
 	
 	Secondly, the $\sigma_z$ term of $h_{\rm eff,\pm}$ may be interpreted as a mass term. In that case, the mass $m$ has to be momentum dependent, which is an exotic and counter-intuitive result. Along this line, $h_{\rm eff,\pm}$ in Eq.~(\ref{nleffh}) can then be rewritten as
   	$h_{\rm eff,\pm}=ck_x\sigma_x+ck_y\sigma_y+m_\pm c^2\sigma_z$ in the same manner as a Dirac particle, with $c=1$, $m= \frac{2\mathcal{B}}{\sqrt{g^2-4\mathcal{B}^2}} k$. Further,
   	the group velocity, \textit{i.e.}, the gradient of the energy band with respect to $k$ near a Dirac point, is given by
    $\frac{g}{\sqrt{g^2-4\mathcal{B}^2}}>1$. It depends on $\mathcal{B}$ and $g$, but stays always larger than $c=1$, indicating a ``superluminal" behavior. Namely, a quasi-particle described by an NDC may travel faster than the effective speed of light in the system. These features can be useful for the quantum simulation of the so-called tachyonic particle~\cite{superlu1,superlu2}. For example, it is of interest to look into the quantum Landau levels and Klein tunnelling of such exotic quasi-particles.

\begin{figure}
		\begin{center}
			\includegraphics[scale=0.45]{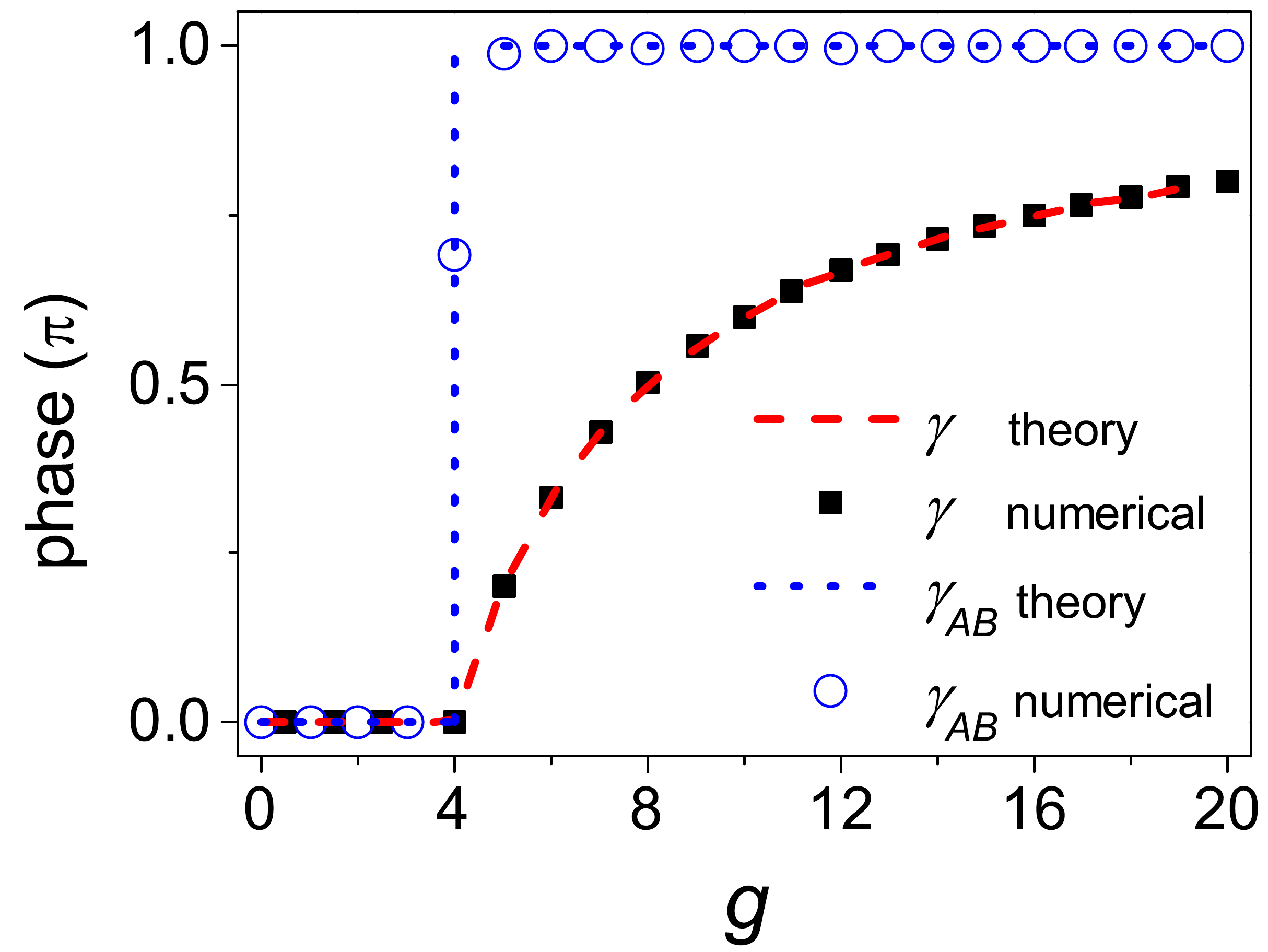}
		\end{center}
		\caption{(color online). (Red) dashed line and black squares depict the Berry phase $\gamma$ associated with one nonlinear Dirac cone, obtained from theory and from direct numerical results based on the time-dependent GP equation. (Blue) line and circles are the adiabatic AB phase $\gamma_{\text{AB}}$  associated with two interfering paths enclosing the same nonlinear Dirac cone, obtained theoretically or computationally. The system parameters chosen are $M=-1$, $B=2$, and $J_x=J_y=1$. The two interfering paths in the momentum space to generate the AB phase are chosen as two small semi-circles going around the nonlinear Dirac cone, one clockwise and the other one counterclockwise. The system is forced to quasi-adiabatically move along the two paths with $\frac{{\rm d}\phi}{{\rm d}t}= 10^{-3}$.
}
		\label{Ber}
	\end{figure}

  	Thirdly, around the self-crossing point of an NDC, an interesting pseudo-spin texture arises. This can be appreciated more clearly by specifically writing down the Bloch band states in the NDC regime. Using $h_{\rm eff,\pm}$ in Eq.~(\ref{nleffh}),  one obtains the following Bloch band pseudo-spinors,
  	\begin{equation}
		|\psi_\pm(k_x,k_y)\rangle = \left[\begin{array}{c} \cos\left(\frac{\theta}{2}\right) \\ \pm\sin\left(\frac{\theta}{2}\right) e^{\mathrm{i}\phi(k_x,k_y)} \end{array}\right]\;,
		\label{gs22}
	\end{equation}
	with
	\begin{eqnarray}
	\tan[\phi(k_x,k_y)]=\frac{k_y}{k_x};\ \tan(\theta)=\frac{\mp\sqrt{g^2-4\mathcal{B}^2}}{2\mathcal{B}}. \label{thetaex}
	\end{eqnarray}
  	Hence, within the NDC regime, the orientation of Bloch state pseudo-spinor is in the $(\theta,\phi)$ direction on the pseudo-spin Bloch sphere, with $\theta$ being independent of $k_x$ and $k_y$. For $g$ close to $g_c=2\mathcal{B}$, the pseudo-spinor is aligned almost towards the north or south pole, and only for $g\gg g_c$, the pseudo-spinor is aligned almost towards the equator. In general situations, the pseudo-spinor may be aligned towards any direction.
	Consider then a parallel transport of the vector $|\psi_\pm(k_x,k_y)\rangle$ around the Dirac cone for one complete cycle~($\phi\rightarrow \phi+2\pi$) (note that the Berry phase itself can be unrelated to any dynamical evolution).
	Both $|\psi_+(k_x,k_y)\rangle$ and $|\psi_-(k_x,k_y)\rangle$ are found to yield the same Berry phase $\gamma$, with
  	\begin{eqnarray}
  		\gamma & = &  i \int \langle \psi_\pm(k_x,k_y) |\frac{{\rm d}}{{\rm d}\phi}|\psi_{\pm}(k_x,k_y)\rangle\ {\rm d}\phi  \nonumber \\
   		&=& \pi W_c \left[1-\cos(\theta)\right].
   		\label{berry}
    \end{eqnarray}
    Here, $W_c=1$ is the winding number of the pseudo-spinor around a string pointing at the north pole. More generally, $W_c=\frac{1}{2\pi} \oint \frac{{\rm d}\phi}{{\rm d}\xi}{\rm d}\xi$, with $\xi$ being an arbitrary parameter of a cyclic path.
    In particular, the quantity $\pi W_c$, the topological part of the Berry phase~\cite{bersc}, is by definition quantized in $\pi$. However, the overall Berry phase $\gamma$ for one NDC is apparently not quantized. Rather, it changes continuously from $\gamma=0$ to $\gamma=\pi$, as $\theta$ changes continuously from $\theta=0$~(when $g\rightarrow g_c$ from above) to $\theta=\pi/2$ (when $g\gg g_c$).
	A computational example without making any approximation is shown in Fig.~\ref{Ber}~(red line and black squares). There, the Berry phase is zero in the absence of NDC. Once an NDC emerges within the area enclosed by a cyclic path in the momentum space, $\gamma$ becomes continuously tunable with $g$. The computational results are in full agreement with our theory. The features of $\gamma$ shown in Fig.~\ref{Ber} also indicate that $\gamma$ can serve as an order parameter to signify the generation of one NDC by interaction.

	\textit{Detection of NDCs and topological invariant.--} A recent study~\cite{ABexp} demonstrated the detection of a conventional $2$D Dirac cone by use of an interference setup, much similar to an AB-effect experiment~\cite{ABphase}. The two interfering paths enclosing a Dirac point are designed in the reciprocal~(quasimomentum) space, with the beam splitting and recombination executed adiabatically by, for example, certain spin-dependent (in our case, sublattice-dependent) force~\cite{ABexp}. We are thus motivated to consider the possibility of detecting NDCs using this interference approach.
	Let the two interfering paths share the same starting point~[see Fig.~\ref{fnlQWZ}(e)] and both go around a Dirac point in a symmetric manner, with one clockwise and the other one counterclockwise. Note that in the NDC regime with all other system parameters fixed, the $\theta$ parameter given in Eq.~(\ref{thetaex}) is a constant along each path. Hence only the $\phi$ parameter [also defined in Eq.~(\ref{thetaex})] suffices to parametrize the two paths. After the system has been forced to adiabatically travel along the two respective paths, we look into their quantum phase difference, called the adiabatic AB phase here.
 	Explicit implementations of how the initial Bloch state is split and recombined, as well as how quasimomenta $k_x$ and $k_y$ are adiabatically varied, are not needed in our computational and theoretical studies below.

 	It is tempting to associate the adiabatic AB phase with the Berry phase $\gamma$ derived in Eq.~(\ref{berry}).  However, computational results clearly indicate that they are different.  As depicted by the blue circles in Fig.~\ref{Ber}, the adiabatic AB phase is found to be zero in the absence of NDCs~($g<g_c$). However, so long as $g>g_c$, the adiabatic AB phase changes discontinuously to $\pi$. That is, contrary to the continuous behavior of $\gamma$ shown above, the adiabatic AB phase is actually quantized in $\pi$ ! For the computational results shown in Fig.~\ref{Ber}, the two paths enclosing an NDC are chosen to be two small semi-circles. We have considered other path geometries and the same results are obtained. The experimentally measurable AB phase obtained here is thus smoking-gun evidence of the formation of an NDC.
 	
	To explain our computational findings, we note that the detection of the AB phase necessarily involves a process of dynamical evolution. Indeed, the system is designed to adiabatically evolve along two different paths. During this process, though $\phi$ changes very slowly at a small rate $\epsilon$, the system cannot be precisely on its instantaneous Bloch states $|\psi(k_x, k_y)\rangle$. Instead, to the first order of $\epsilon$, the system's actual time-evolving state $|\Psi(t)\rangle$ is given by
	\begin{equation}
		|\Psi(t)\rangle=  |\psi(k_x,k_y)\rangle + \epsilon |\delta \psi (t)\rangle \;.
	\end{equation}
 	This adiabatic perturbation to the first order of $\epsilon$ has a nontrivial impact on the nonlinear Hamiltonian $\mathcal{H}[k_x,k_y,\Psi(t)]$ governing the dynamics~\cite{liujie,gongAOP}, resulting in
	\begin{equation}
		\mathcal{H}[k_x,k_y,\Psi(t)]=\mathcal{H}[k_x,k_y,\psi(k_x,k_y)] + h_{\text{NA}}+ O(\epsilon^2)\;,
		\label{expand}
	\end{equation}
	where $h_{\text{NA}}$ stands for the first-order nonadiabatic contribution to system's instantaneous effective Hamiltonian $\mathcal{H}[k_x,k_y,\psi(k_x,k_y)]$.  Using the first-order perturbation theory and Eq.~(\ref{gs22}), we find that $h_{\text{NA}}$  is rather simple \cite{supp}:
	\begin{eqnarray}
		h_{\text{NA}}=\frac{1}{2} \sigma_z \frac{{\rm d}\phi(k_x,k_y)}{{\rm d}t}.   \label{hna}
		\label{Ena}
	\end{eqnarray}
	\n Note that the first-order energy correction due to $h_{\text{NA}}$, i.e., $\mathcal{E}_{\rm NA}=\langle \psi(k_x,k_y)| h_{\rm NA}|\psi(k_x,k_y)\rangle$, is simply given by $\frac{1}{2} \cos(\theta)\frac{{\rm d}\phi(k_x,k_y)}{{\rm d}t}$.
As a useful observation, the two paths now yield different $\mathcal{E}_{\rm NA}$ as energy corrections because their $\frac{{\rm d}\phi(k_x,k_y)}{{\rm d}t}$ has different signs.

	Consider then the dynamical phase along each path, which is obtained as an integral of $E_{\pm}(k_x,k_y)+\mathcal{E}_{\rm NA}$, i.e.,
the expectation value of $\mathcal{H}[k_x,k_y,\Psi(t)]=\mathcal{H}[k_x,k_y,\psi(k_x,k_y)]+h_{\text{NA}}$,
 over a time scale of the order $\epsilon^{-1}$. Clearly then,
	though $\mathcal{E}_{\rm NA}$ is of the order of $\epsilon$, with its impact accumulated over time of the order $\epsilon^{-1}$,
 	it can still make a contribution of the order of $\epsilon^{0}$ to the dynamical phase.
	Further, we note  that the dynamical phase contributed by $E_{\pm}(k_x,k_y)$ is identical for two symmetric interfering paths on the same branch of an NDC. Thus,
	only $\mathcal{E}_{\rm NA}$ introduces an $\epsilon$-\textit{independent} dynamical phase difference $\Delta \gamma_{\text{dyn}}$ between the two adiabatic paths, with
	\begin{eqnarray}
		\Delta \gamma_{\text{dyn}} &=& \frac{1}{2}\cos(\theta) \left [ \int^{\pi}_0 {\rm d}\phi - \int_{0}^{-\pi}{\rm d}\phi  \right] \nonumber \\
 		&=& \pi \cos(\theta).
	\end{eqnarray}
	The adiabatic AB phase produced by the two interfering paths is then given by $\Delta \gamma_{\text{dyn}}$ obtained above {\it plus} their geometric phase difference. Recognizing that their geometric phase difference is simply the previously found Berry phase $\gamma$~(dropping a negligible correction of the order of $\epsilon$), we obtain the adiabatic AB phase as the following,
	\begin{eqnarray}
		\gamma_{\text{AB}}=\gamma+ \Delta \gamma_{\text{dyn}}= \pi W_c.
	\end{eqnarray}
	That is, $\Delta \gamma_{\text{dyn}}$ precisely cancels the non-topological part of the Berry phase $\gamma$, yielding an adiabatic AB phase quantized in $\pi$ !  This fully explains the computational results presented in Fig.~\ref{Ber}~(see blue dashed line and circles). Our theory here also identifies the winding number $W_c$ as an observable topological invariant of NDCs. In our numerically exact calculations, we also find that if the rate of change in $\phi$ is increased by ten times from what is considered in Fig.~\ref{Ber}, the quantization of the AB phase is only slightly degraded due to nonadiabatic effects beyond the first order of $\epsilon$.

	\textit{Discussion and conclusions.--} It is also of interest to discuss the role of the system parameter $M$. In the non-interacting Dirac-Chern insulator model, cases with $|M|>2$~($|M|<2$) represent a  topologically trivial~(nontrivial) phase~\cite{QWZ}, with a topological phase transition at $M=2$. The same behavior is observed in the interacting case with $g<g_c$. Interestingly, with the emergence of NDCs~[$g>g_c=2B(M+2)$], regardless of the value of $M$, their topological invariant $W_c$ stays to be unity and the adiabatic AB phase remains quantized in $\pi$. In particular, for $g>8B$, phases with $|M|>2$ and those with $|M|<2$ all have the same topological invariant
 	and hence they can be categorized into the same~(gapless) topological phase. This clearly shows the possibility of two topologically distinct phases to become topologically equivalent as an outcome of interaction~\cite{inttp1}, thus revealing the interplay of topology and nonlinearity. About possible experimental studies, one potential issue is the dynamical stability \cite{loop3} of our nonlinear two-band model. We have checked that near a looped subband, the lower main band is dynamically stable in the presence of perturbations that respect the translational symmetry of the system (see Supplemental Material for the calculation detail).
 	
	In conclusion, we have discovered a nonlinearity-induced quantum phase transition from
	a known topological insulating phase to a novel gapless topological phase featured by NDCs. The NDCs
  	are robust against local perturbations without symmetry restrictions. In addition, they have peculiar pseudo-spin textures, remarkable band structures, and non-quantized Berry phases. By showing that an adiabatic AB phase is still quantized in $\pi$, we have identified a winding number as a directly measurable topological invariant of such nonlinear Dirac cones.
  	
  	\vspace{1cm}
  	
  	\n {\bf Acknowledgements}
  	
  	J.G. is supported by the Singapore NRF grant No. NRF-NRFI2017-04 (WBS No. R-144-000-378-281) and by the Singapore Ministry of Education Academic Research Fund Tier I (WBS No. R-144-000-353-112).  W.Z. is supported by the National Natural Science Foundation of China (Grant No. 11447016), as well as the Foundation of China Scholarship of Council (Grant No. 201508360124).
  	
  	\appendix
  	
  	\vspace{1cm} \begin{center} {\bf Supplemental Material} \end{center}
  	
  	This supplementary material has seven sections. In Appendix~A, we describe a derivation of Eq.~(2) in the main text, namely, the GP Hamiltonian in the momentum space. In Appendix~B, we show how to derive the energy dispersion relation and its associated effective Hamiltonian near the self-crossing point of a nonlinear Dirac cone~(NDC) at $k_x=k_y=0$. In Appendix~C, we present computational examples to confirm the robustness of NDCs against local perturbations along any direction. In Appendix~D, we derive $h_{\text{NA}}$, namely, a nonadiabatic correction to the instantaneous GP Hamiltonian.  In Appendix~E, to help readers to reproduce our results in Fig.~2 in the main text, we present some computational details. In Appendix~F, we elucidate how our model arises as a mean-field approximation of interacting many-body bosons. Finally, in Appendix~G, we present the detailed calculation of dynamical stability against perturbations that respect the translational symmetry of the system.

\section{Lattice Hamiltonian and Nonlinear Bloch Bands}

Under the tight-binding approximation, the stationary Schr\"{o}dinger equation
for the nonlinear Dirac-Chern insulator model is described by the following:
\begin{eqnarray}
	E \phi_{s,i,j} &=& s B M \phi_{s,i,j} +s \frac{B}{2} \left(\phi_{s,i+1,j}+\phi_{s,i-1,j}+\phi_{s,i,j+1} +\phi_{s,i,j-1}\right) \nonumber \\
	&& +\frac{J_x}{2\mathrm{i}} \left(\phi_{-s,i+1,j}-\phi_{-s,i-1,j}\right) -s\frac{J_y}{2} \left(\phi_{-s,i,j+1}-\phi_{-s,i,j-1}\right) + g |\phi_{s,i,j}|^2 \phi_{s,i,j} \;,
	\label{lattice}
\end{eqnarray}
\n where $s=\pm 1$ stands for the pseudo-spin index, $i$ and $j$ indices denote the lattice sites along $x$ and $y$ directions, respectively. Under periodic boundary conditions, the Bloch theorem can be applied to write $\phi_{+,i,j}=e^{\mathrm{i}(k_x i+k_y j)}\psi_1$ and $\phi_{-,i,j}=e^{\mathrm{i}(k_x i+k_y j)}\psi_2$. This yields Eqs.~(1) and (2) in the main text.

\section{Energy dispersion and effective Hamiltonian of the Nonlinear Dirac Cone}

Generically, a solution to Eq.~(1) in the main text satisfies the following conditions,

\begin{eqnarray}
	|\psi_1|^2 &=& \frac{1}{2}+\frac{\mathcal{B}}{2(E-g)} \; \label{proper2} \\
	|\psi_2|^2 &=& \frac{1}{2}-\frac{\mathcal{B}}{2(E-g)} \;,\label{proper3} \\
	0 &=& E^4-3gE^3+E^2\left(\frac{13}{4}g^2-\mathcal{B}^2-|\gamma|^2\right)+E\left(g\mathcal{B}^2+2g|\gamma|^2-\frac{3}{2}g^3\right)  \nonumber \\
	&& +\frac{1}{4}g^4-\frac{1}{4}g^2\mathcal{B}^2-g^2|\gamma|^2 \;.
	\label{proper}
\end{eqnarray}

\n When $k_x=k_y=0$, Eq.~(1) in the main text can be immediately solved, which gives two real solutions $E_{1}=g+ \mathcal{B}$ and $E_{2}=g- \mathcal{B}$ for small $|g|<2 |\mathcal{B}|$. As the nonlinear strength $|g|$ is increased beyond $2|\mathcal{B}|$, an additional real solution with $E_0=\frac{g}{2}$ emerges, which is doubly degenerate with eigenvectors given by

\begin{eqnarray}
	|\Psi_1 \rangle &=& \frac{1}{\sqrt{2}}\left(\begin{array}{c}\sqrt{1-\frac{2\mathcal{B}}{g}} \\ \sqrt{1+\frac{2\mathcal{B}}{g}} \end{array}\right) \;, \\
	|\Psi_2 \rangle &=& \frac{1}{\sqrt{2}}\left(\begin{array}{c}\sqrt{1-\frac{2\mathcal{B}}{g}} \\ -\sqrt{1+\frac{2\mathcal{B}}{g}} \end{array}\right) \;.
\end{eqnarray}

\n In particular, $E_0$ corresponds to a self-crossing point in the energy dispersion and is responsible for the development of a loop structure in Fig.~1(c) in the main text. At small $k_x$ and $k_y$, an energy solution can be obtained perturbatively according to

\begin{equation}
	E=E^{(0)}+E^{(1)}+\cdots \;,
\end{equation}

\n where $E^{(0)}$ is an energy solution at $k_x=k_y=0$, and $E^{(n)}$ represents a component of $E$ that is of the order of $k_x^n$ or $k_y^n$. By expanding Eq.~(\ref{proper}) up to the first order in $k_x$ and $k_y$, we obtain

\begin{equation}
	0 = \left\{4\left[E^{(0)}\right]^3-9g\left[E^{(0)}\right]^2+2E^{(0)}\left(\frac{13}{4}g^2-\mathcal{B}^2\right)+\left(g\mathcal{B}^2-\frac{3}{2}g^3\right)\right\}E^{(1)}\;.
	\label{1st}
\end{equation}

\n Since we are interested in finding the energy dispersion near an NDC, we take $E^{(0)}=E_0=\frac{g}{2}$. It can be verified that the terms inside the curly bracket add up to $0$, implying that $E^{(1)}$ is not necessarily $0$. To find $E^{(1)}$, we expand Eq.~(\ref{proper}) up to the second order in $k_x$ and $k_y$, which leads to

\begin{eqnarray}
	0 &=& \left[E^{(1)}\right]^2 \left(\frac{1}{4}g^2-\mathcal{B}^2\right)-\left(\frac{g}{2}\right)^2(k_x^2+k_y^2); \nonumber \\
	E^{(1)} &=& \pm \frac{1}{\sqrt{1-\frac{4\mathcal{B}^2}{g^2}}}\sqrt{k_x^2+k_y^2} \;.
\end{eqnarray}

\n Up to the first order in $k_x$ and $k_y$, energy dispersion near a self-crossing at $k_x=k_y=0$ is thus given by Eq.~($3$) in the main text.


Substituting Eq.~(3) in the main text into Eqs.~(\ref{proper2}) and (\ref{proper3}) here, we obtain

\begin{eqnarray}
	|\psi_1|^2 &\approx & \frac{1}{2} -\frac{\mathcal{B}}{g} \left(1\pm \frac{1}{\sqrt{g^2-4\mathcal{B}^2}}\sqrt{k_x^2+k_y^2}\right) \;, \label{proper4} \\
	|\psi_2|^2 &\approx & \frac{1}{2} +\frac{\mathcal{B}}{g} \left(1\pm \frac{1}{\sqrt{g^2-4\mathcal{B}^2}}\sqrt{k_x^2+k_y^2}\right) \;. \label{proper5}
\end{eqnarray}

\n Eqs.~(\ref{proper4}), (\ref{proper5}), and Eq.~($3$) in the main text then allow us to recast Eq.~($1$) in the main text in the form

\begin{equation}
	\left(\begin{array}{cc} \mp \frac{2\mathcal{B}}{g\sqrt{1-\frac{4\mathcal{B}^2}{g^2}}}\sqrt{k_x^2+k_y^2} & k_x-\mathrm{i}k_y	\\ k_x+\mathrm{i}k_y & \pm \frac{2\mathcal{B}}{g\sqrt{1-\frac{4\mathcal{B}^2}{g^2}}}\sqrt{k_x^2+k_y^2} \end{array}\right) \left( \begin{array}{c} \psi_1 \\ \psi_2 \end{array}\right) = \pm\frac{1}{\sqrt{1-\frac{4\mathcal{B}^2}{g^2}}}\sqrt{k_x^2+k_y^2}\left( \begin{array}{c} \psi_1 \\ \psi_2 \end{array}\right)  \;.
	\label{states}
\end{equation}

\n The left hand side of Eq.~(\ref{states}) can be written in terms of Pauli matrices to yield $h_{\rm eff,\pm}$ as given in the main text. It then follows that Eq.~(\ref{states}) takes the form of an eigenvalue equation associated with the positive~(negative) energy solution of an effective Hamiltonian $h_{\rm eff,-}$ ($h_{\rm eff,+}$). This thus shows that the positive and negative energy branches of an NDC are described by two different effective Hamiltonians.

\section{On the robustness of NDCs against local perturbations}
The effective Hamiltonian $h_{\rm eff,\pm}$ derived above suggests that NDCs we find are robust to perturbations along all directions (hence no symmetry requirements). This is in contrast to conventional 2D Dirac cones that can be easily destroyed by a perturbative mass term.
To check this understanding, Fig. 1 below
depicts how various local perturbations of the form $h_x\sigma_x+h_y\sigma_y+h_z\sigma_z$ affect an NDC. For better view, the perturbed but surviving Dirac cones are projected onto one value of $k_x$.
\begin{figure}
	\begin{center}
		\includegraphics[scale=0.40]{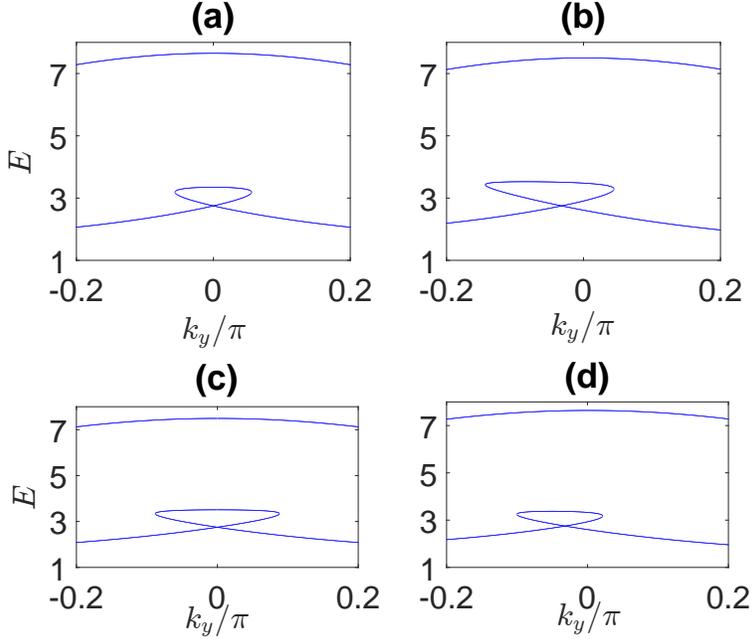}
	\end{center}
	
	\caption{A self-crossing nonlinear Dirac cone survives after switching on various local perturbations of the form $h_x\sigma_x+h_y\sigma_y+h_z\sigma_z$. Parameters chosen are $M=-1$, $B=2$, $J_x=J_y=1$,  and  $h_x=h_y=0$, $h_z=0.15$ in (a), $h_x=h_z=0$, $h_y=0.1$ in (b), $h_y=h_z=0$,  $h_x=0.1$ in (c), and $h_x=h_y=0.1$, $h_z=0.15$ in (d).  We choose $k_x=0$ in (a) and (b) and $k_x=-\arcsin\left(0.1\right)$ in (c) and (d).}
	\label{perturb}
\end{figure}

\section{Nonadiabatic correction to the GP Hamiltonian}

We start by separating the phase containing explicit time dependence from a wave function in the projective Hilbert space. Specifically, we  introduce $\Phi_{1(2)}=e^{-\mathrm{i}f}\Psi_{1(2)}$, where $\Psi_{1(2)}$ is one pseudo-spinor component of an actual time-evolving state, $f$ contains both dynamical and geometrical phases attained by $\Psi_{1(2)}$ as a function of time, and $\Phi_{1(2)}$ represents a pseudo-spinor component of a state in the projective Hilbert space. While $\Phi_{1(2)}$ does not have explicit time dependence, they may still depend on time implicitly through their dependence on time-dependent system parameters. The time-dependent GP equation can then be written as

\begin{equation}
	\frac{{\rm d}f}{{\rm d}t} \left( \begin{array}{c} \Phi_1 \\ \Phi_2 \end{array}\right) = \mathrm{i} \frac{\partial}{\partial t}\left( \begin{array}{c} \Phi_1 \\ \Phi_2 \end{array}\right)-\mathcal{H}[k_x,k_y,\Phi(k_x,k_y)]\left( \begin{array}{c} \Phi_1 \\ \Phi_2 \end{array}\right)\;,
	\label{prep1}
\end{equation}

\n where $\mathcal{H}[k_x,k_y,\Phi(k_x,k_y)]$ as a state-dependent Hamiltonian is defined by Eq.~($2$) in the main text and $|\Phi\rangle =\left(\Phi_1, \Phi_2\right)^T$. Upon multiplying Eq.~(\ref{prep1}) from the left by $\left(\Phi_1^*, \Phi_2^*\right)$ and using normalization condition, we arrive at

\begin{equation}
	\frac{{\rm d}f}{{\rm d}t} = \sum_{a=1,2}\mathrm{i}\Phi_a^*\frac{\partial}{\partial t}\Phi_a -\sum_{a,b=1,2} \Phi_a^* \mathcal{H}_{ab} \Phi_b \;.
	\label{prep2}
\end{equation}

Next, we assume that $k_x$ and $k_y$ are adiabatically tuned in time along a closed loop in the 2D BZ enclosing a self crossing point, such that $\frac{{\rm d}\mathcal{H}}{{\rm d}t}\propto \epsilon\ll 1$. By expanding $\frac{{\rm d}f}{{\rm d}t}$ and $\Phi_{1(2)}$ perturbatively in $\epsilon$, one has

\begin{eqnarray}
	\frac{{\rm d}f}{{\rm d}t} &=& \frac{{\rm d}f^{(0)}}{{\rm d}t}+\epsilon \frac{{\rm d}f^{(1)}}{{\rm d}t}+\cdots \;, \nonumber \\
	\Phi_a &=& \Phi_a^{(0)}+\epsilon \Phi_a^{(1)}+\cdots \;,
	\label{pert}
\end{eqnarray}

\n where $a=1,2$. Without loss of generality, we assume that the state is initially prepared in the stationary state $|\Phi(t=0)\rangle=|\Phi_0(k_{x,0},k_{y,0})\rangle $ of $\mathcal{H}$ associated with energy $E_+$ given by Eq.~($3$) in the main text, where $k_{x,0}$ and $k_{y,0}$ are the quasimomenta at time $t=0$. The zeroth order state at any time $t$ is then given by the instantaneous stationary solution to $\mathcal{H}$ associated with the same energy band, \textit{i.e}., $|\Phi^{(0)}(t)\rangle =|\Phi_0[k_x(t),k_y(t)]\rangle $. Equation~(\ref{prep2}) can then be solved perturbatively as

\begin{eqnarray}
	\frac{{\rm d}f^{(0)}}{{\rm d}t}&=& -E_{+}[k_x(t),k_y(t)]\;, \nonumber \\
	\epsilon\frac{{\rm d}f^{(1)}}{{\rm d}t}&=&\mathrm{i}\langle \Phi_0 |\frac{\partial}{\partial t}|\Phi_0\rangle -2\epsilon g\sum_{a=1,2} \mathrm{Re}\left[(\Phi_{0,a}^*)^2\Phi_a^{(1)}\Phi_{0,a}\right]\;.
	\label{pert2}
\end{eqnarray}

\n It is noted that upon integration over one period of adiabatic evolution, $f^{(0)}$ is the usual dynamical phase as defined in linear systems, the first term of $\epsilon f^{(1)}$ corresponds to Berry phase as also defined in linear systems, while the second term is an extra term induced by nonlinearity. By plugging Eq.~(\ref{pert2}) into Eq.~(\ref{prep1}) and keep only terms proportional to $\epsilon^1$, we arrive at

\begin{eqnarray}
	\mathrm{i} \sum_{b=1,2} \left( \Phi_{0,b}^* \Phi_{0,a}-\delta_{ab} \right)\frac{\partial}{\partial t}\Phi_{0,b}&=& \epsilon\sum_b \left(E_+\delta_{ab}-\mathcal{H}_{ab}-g|\Phi_{0,a}|^2\delta_{ab}\right) \Phi_b^{(1)} \nonumber \\
	&& +2\epsilon g \Phi_{0,a}\sum_b \mathrm{Re}\left[(\Phi_{0,b}^*)^2\Phi_b^{(1)}\Phi_{0,b}\right]-\epsilon g (\Phi_{0,a})^2[\Phi_a^{(1)}]^*\;,
	\label{check1}
\end{eqnarray}

\n where $a=1,2$ and normalization of $\Phi_a$ up to the first order of $\epsilon$ leads to

\begin{equation}
	\sum_{a=1,2} \mathrm{Re}\left[\Phi_{0,a}^*\Phi_a^{(1)}\right]=0\;.
	\label{nor}
\end{equation}

\n We now write $\Phi_{0,a}$ in the form

\begin{equation}
	\left( \begin{array}{c} \Phi_{0,1} \\ \Phi_{0,2} \end{array}\right) =   \left[ \begin{array}{c} \cos\left(\frac{\theta}{2}\right) \\ \sin\left(\frac{\theta}{2}\right)e^{\mathrm{i}\phi} \end{array}\right]\;.
	\label{gs}
\end{equation}

\n By taking $a=1$ and considering only the real part of Eq.~(\ref{check1}) as well as Eq.~(\ref{nor}), up to zeroth order terms in $k_x$ and $k_y$ we obtain

\begin{equation}
	\epsilon \mathrm{Re}\left[\Phi_{0,2}^*\Phi_2^{(1)}\right] = - \frac{1}{4g} \frac{{\rm d}\phi}{{\rm d}t} \;.
	\label{reals}
\end{equation}

\n As a last step, at a given time $t$ we expand $\mathcal{H}[k_x,k_y,\Phi(t)]$ explicitly up to the first order in $\epsilon$,

\begin{eqnarray}
	\mathcal{H}[k_x,k_y,\Phi(t)] &=& \left(\begin{array}{cc} \mathcal{B}(k_x,k_y) + g \Big|\Phi_{0,1}+\epsilon \Phi_1^{(1)}\Big|^2 & \gamma(k_x,k_y)	\\ \gamma^*(k_x,k_y) & -\mathcal{B}(k_x,k_y)+g\Big|\Phi_{0,2}+\epsilon \Phi_2^{(1)}\Big|^2 \end{array}\right)+\mathcal{O}(\epsilon^2) \nonumber \\
	&=& \mathcal{H}(k_x,k_y,\Phi_0)+\left(\begin{array}{cc} 2\epsilon g \mathrm{Re}\left[\Phi_{0,1}^*\Phi_1^{(1)}\right] & 0	\\ 0 & 2\epsilon g \mathrm{Re}\left[\Phi_{0,2}^*\Phi_2^{(1)}\right] \end{array}\right)+\mathcal{O}(\epsilon^2) \nonumber \\
	&=& 	\mathcal{H}(k_x,k_y,\Phi_0)+ \frac{1}{2} \frac{d\phi}{dt} \sigma_z \;,
	\label{expand2}
\end{eqnarray}

\n where we have used Eqs.~(\ref{reals}) and (\ref{nor}) to arrive at the last line. Upon comparing Eq.~(\ref{expand2}) with Eq.~($9$) in the main text, we identify $h_{\rm NA}$ as

\begin{equation}
	h_{\rm NA} =\frac{1}{2} \frac{{\rm d}\phi}{{\rm d}t} \sigma_z \;.
\end{equation}

\section{Computational studies of a proposed AB-effect experiment}

At $(k_x,k_y)=(-R,0)$, we first solve Eq.~($1$) in the main text numerically and find its stationary solutions for the lower branch of the NDC. This state is chosen as the initial condition and then evolved adiabatically following clockwise or counterclockwise semi-circle (with a radius $R=10^{-4}$) in the $k$-space, as sketched in Fig.~$1$(e) in the main text. The full circle is parametrized as $k_x(t)=R\cos[\phi(t)], k_y(t)=R\sin[\phi(t)]$, with $\phi(t)$ being the slowly varying parameter in time $t$. Numerically, the actual time-evolving states are obtained by use of the splitting operator method. To that end, we decompose the propagator for one adiabatic path as $U[(-R,0)\rightarrow(R,0)]=e^{-i{\cal H}_N{\rm d}t}\cdots e^{-i{\cal H}_2{\rm d}t}e^{-i{\cal H}_1{\rm d}t}$, where ${\cal H}_n={\cal H}\{k_x(n\cdot{\rm d}t),k_y(n\cdot{\rm d}t),\psi[k_x(n\cdot{\rm d}t),k_y(n\cdot{\rm d}t)]\}$, and ${\rm d}t=T/N$ is a sufficiently small time interval. In our simulation, we choose the total time of adiabatic evolution $T=2\pi\times10^3$ and $N=10^6$, so as to achieve a
nearly adiabatic evolution and a high numerical precision of the splitting operator method.  The evolution ends at $(k_x,k_y)=(R,0)$, where the phase difference of the two states accumulated following the two symmetric paths is found by finding the phase angle of their overlap. This relative phase determines the location of the central interference pattern of the adiabatic AB-effect associated with the NDC. As  a comparison, the Berry phase can be computationally found in a similar fashion, by finding the difference between the overall phase acquired by a time evolving state
(after adiabatically moving around the NDC for one complete cycle) and the accumulated dynamical phase $\int \langle\Psi(t)|{\cal H}(t)|\Psi(t)\rangle {\rm d}t$ (That is, integration of the expectation value of ${\cal H}(t)$ evaluated on the actual time evolving state $|\Psi(t)\rangle$  during the whole adiabatic cycle).

\section{Nonlinear model as a mean-field approximation of interacting many-body bosons.}

Consider an interacting many-body bosonic system described by the following second quantized Hamiltonian

\begin{eqnarray}
	H &=& \sum_{i,j} \left\lbrace \frac{BM}{2} \hat{\Psi}_{i,j}^\dagger \sigma_z \hat{\Psi}_{i,j}+\frac{B}{2} \left[2\hat{\Psi}_{i,j}^\dagger \sigma_z \hat{\Psi}_{i+1,j}+\hat{\Psi}_{i,j}^\dagger \sigma_z \hat{\Psi}_{i,j+1}\right]\right.\nonumber \\
	&& \left. \frac{J_x}{2\mathrm{i}}\hat{\Psi}_{i,j}^\dagger \sigma_x \hat{\Psi}_{i+1,j}+\frac{J_y}{2\mathrm{i}}\hat{\Psi}_{i,j}^\dagger \sigma_y \hat{\Psi}_{i,j+1}+h.c.\right\rbrace + \sum_{i,j,s} \frac{g}{2} \hat{\Psi}_{s,i,j}^\dagger \hat{\Psi}_{s,i,j}^\dagger \hat{\Psi}_{s,i,j} \hat{\Psi}_{s,i,j}\;,
	\label{full0}
\end{eqnarray}

\n where $\hat{\Psi}_{i,j}\equiv [\hat{\Psi}_{1,i,j}, \hat{\Psi}_{-1,i,j}]^{T}$, and $\hat{\Psi}_{1(-1),i,j}$ ($\hat{\Psi}_{1(-1),i,j}^\dagger$) is the bosonic annihilation (creation) operator at lattice site $(i,j)$ with pseudo-spin index $1(-1)$. By employing Heisenberg equation,

\begin{eqnarray}
	\mathrm{i}\frac{\partial}{\partial t}\hat{\Psi}_{s,i,j}&=& [\hat{\Psi}_{s,i,j},H] \nonumber \\
	&=& s B M \hat{\Psi}_{s,i,j} +s \frac{B}{2} \left(\hat{\Psi}_{s,i+1,j}+\hat{\Psi}_{s,i-1,j}+\hat{\Psi}_{s,i,j+1} +\hat{\Psi}_{s,i,j-1}\right) \nonumber \\
	&& +\frac{J_x}{2\mathrm{i}} \left(\hat{\Psi}_{-s,i+1,j}-\hat{\Psi}_{-s,i-1,j}\right) -s\frac{J_y}{2} \left(\hat{\Psi}_{-s,i,j+1}-\hat{\Psi}_{-s,i,j-1}\right) + g \hat{\Psi}_{s,i,j}^\dagger \hat{\Psi}_{s,i,j} \hat{\Psi}_{s,i,j}\;.
	\label{full}
\end{eqnarray}

\n Mean-field approximation amounts to replacing $\hat{\Psi}_{s,i,j}$ ($\hat{\Psi}_{s,i,j}^\dagger$) by a classical field $\phi_{s,i,j}=\langle \hat{\Psi}_{s,i,j}\rangle $ ($\phi_{s,i,j}^*$), where the average is taken over the ground state of Eq.~(\ref{full0}). Equation~(\ref{full}) then becomes a self-consistent equation for the order parameter $\phi_{s,i,j}$, which now becomes nonlinear. The stationary solutions for $\phi_{s,i,j}$ can thus be found by solving Eq.~(\ref{lattice}), leading to the energy bands we found in this work. The mean-field approximation above corresponds to retaining only the wavefunction of the condensate, while thermal and quantum fluctuations are completely neglected \cite{BECbook}. Nevertheless, such a mean field theory has been successful to interpret a variety of experimental phenomena \cite{BECbook,reff1,reff2,BECbook2,BECbook3}.
	
	\section{Dynamical stability analysis of the nonlinear bands}
	
	For simplicity, we will focus on perturbations to the stationary solutions that respect the translational symmetry of the system, thus allowing us to use the time dependent GP equation in the momentum space as given by
	
	\begin{equation}
		\mathrm{i}\frac{\partial}{\partial t} |\Psi(k_x,k_y,t)\rangle = \mathcal{H}[k_x,k_y,\Psi(k_x,k_y,t)]|\Psi(k_x,k_y,t)\rangle \;,
		\label{timeGP}
	\end{equation}
	
	\noindent where $\mathcal{H}$ is given by Eq.~($2$) in the main text, to analyze their time evolution. Such perturbations can be generally written in pseudo-spinor components as
	
	\begin{equation}
		|\delta\psi(k_x,k_y,t)\rangle=\left[\begin{array}{c}\delta \psi_1(k_x,k_y,t) \\ \delta \psi_2(k_x,k_y,t) \end{array}\right] \;.
	\end{equation}
	
	Consider now a stationary solution $|\psi(k_x,k_y,t)\rangle$ satisfying Eq.~(1) in the main text with energy $E(k_x,k_y)$. Our objective is to evaluate the time evolution of a state initially prepared near such a stationary solution as given by $|\Psi(k_x,k_y,t)\rangle=|\psi(k_x,k_y,t)\rangle+|\delta \psi(k_x,k_y,t)\rangle$. If $|\Psi(k_x,k_y,t)\rangle$ does not go to $\infty$ as $t\rightarrow \infty$, then the stationary solution $|\psi(k_x,k_y,t)\rangle$ is dynamically stable. For neater analysis, we may separate the ``dynamical phase" from $|\Psi(k_x,k_y,t)\rangle $ as
	
	\begin{eqnarray}
		|\Psi(k_x,k_y,t)\rangle &=& e^{-\mathrm{i}E t} |\Phi(k_x,k_y,t)\rangle \;, \\
		|\psi(k_x,k_y,t)\rangle &=& e^{-\mathrm{i} E t} |\psi(k_x,k_y,0)\rangle \;, \\
		|\delta\psi(k_x,k_y,t)\rangle &=& e^{-\mathrm{i} E t} |\delta \phi(k_x,k_y,t)\rangle \;.
	\end{eqnarray}
	
	\noindent After some algebra, Eq.~(\ref{timeGP}) and its conjugate can be recast in the following form,
	
	\begin{equation}
		\mathrm{i} \frac{\partial}{\partial t} \left(\begin{array}{c}  \delta \phi_1 \\  \delta \phi_2 \\ \delta \phi_1^* \\ \delta \phi_2^*\end{array}\right) = \mathcal{L} \left(\begin{array}{c}  \delta \phi_1 \\  \delta \phi_2\\ \delta \phi_1^* \\ \delta \phi_2^*\end{array}\right) \;,
		\label{stab}
	\end{equation}
	
	\noindent where
	
	\begin{eqnarray}
		\mathcal{L} &=& \left[\begin{array}{cc}  H_{gp}+g\left(\begin{array}{cc}  |\psi_1(k_x,k_y,0)|^2 & 0\\ 0 & |\psi_2(k_x,k_y,0)|^2 \end{array}\right) & g\left(\begin{array}{cc}  \psi_1(k_x,k_y,0)^2 & 0\\ 0 & \psi_2(k_x,k_y,0)^2 \end{array}\right)\\ -g\left(\begin{array}{cc}  \psi^*_1(k_x,k_y,0)^{2} & 0\\ 0 & \psi^*_2(k_x,k_y,0)^{2} \end{array}\right) & -H_{gp}^*-g\left(\begin{array}{cc}  |\psi_1(k_x,k_y,0)|^2 & 0\\ 0 & |\psi_2(k_x,k_y,0)|^2 \end{array}\right) \end{array}\right]\;, \nonumber \\
		&& \\
		H_{gp} &=& \mathcal{H}[k_x,k_y,\psi(k_x,k_y,0)] -E \mathcal{I}_2\;.
	\end{eqnarray}
	
	Due to the resemblance of Eq.~(\ref{stab}) with the time dependent Schr\"{o}dinger equation in linear quantum mechanics, the time evolution of the perturbation is governed by the operator $e^{-\mathrm{i}\mathcal{L} t}$. However, since $\mathcal{L}$ is not a Hermitian operator, eigenvalues of $\mathcal{L}$ can in general be complex. It follows that in order for $|\Phi(k_x,k_y,t)\rangle$ to be dynamically stable, all eigenvalues $\lambda_n$ of $\mathcal{L}$ must satisfy \cite{Yvan},
	
	\begin{equation}
		\mathrm{Im}(\lambda_n) =  0 \, \, \, \, \, \, \, \, \, \, \, \forall n\;.
	\end{equation}
	
	\begin{figure}
		\begin{center}
			\includegraphics[scale=0.40]{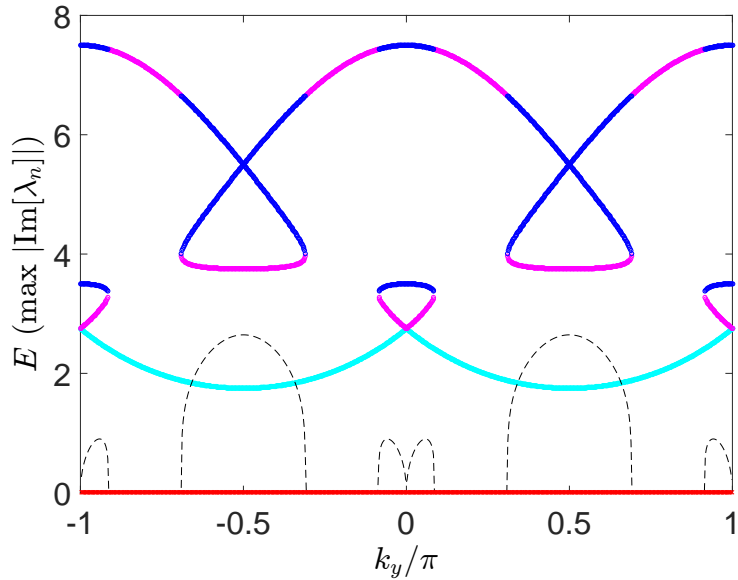}
		\end{center}
		
		\caption{Energy spectrum of the nonlinear QWZ and
			$\text{max}|\text{Im}(\lambda_n)|$ for the first two bands as a function of $k_y$. Parameters
			chosen are $k_x = 0$, $B = 2$, $M = -1$, and $g = 5.5$. The red
			crossed line corresponds to $\text{max}|\text{Im}(\lambda_n)|$ associated with the first
			band (cyan colored band), whereas the black dashed line corresponds
			to $\text{max}|\text{Im}(\lambda_n)|$ associated with the second band (magenta
			colored band).}
		\label{stabs}
	\end{figure}
	
	Figure~\ref{stabs} shows the maximum imaginary component of all the eigenvalues of $\mathcal{L}$ for the first two bands. It follows that the lowest energy band is dynamically stable throughout the $k_y$ BZ, while the second band is dynamically unstable. Though not shown in the figure, we have also checked that the lowest band is also dynamically stable for other values of $k_x$.



\begin{thebibliography}{99}
\bibitem{footnote} Note that a perturbation $h_z\sigma_z$ will not simply appear as an additional term in the effective Hamiltonian. Instead, we need to return to Eq.~(\ref{GPeq1}) and retrace our derivation leading to Eq.~(\ref{nleffh}). By inspection, such a perturbative mass term will simply combine with $\mathcal{B}$ to become $\mathcal{B}'=\mathcal{B}+h_z$. Consequently, Eq.~(\ref{nleffh}) will retain its form, with $\mathcal{B}$ being replaced by $\mathcal{B}'$.
\end{thebibliography}
\end{document}